%% file: main.tex
\documentclass[12pt]{article}

\usepackage[T1]{fontenc}
\usepackage[utf8]{inputenc}
\usepackage{amsmath,amssymb}
\usepackage{hyperref}
\usepackage{tikz}
\usepackage[margin=1in]{geometry}
\usepackage{natbib}

\title{The Fine-Structure Constant as a Scaled Quantity}
\author{Harry Sticker\\
  \small Ganymede Technology, New York, New York, United States\\
  \small \texttt{hsticker@ganymedetechnology.com}}
\date{}

\begin{document}

\maketitle

\begin{abstract}
The fine-structure constant $\alpha \approx 1/137$ is traditionally regarded as a fundamental dimensionless parameter. I argue instead that $\alpha$ is a scaled quantity that arises only where the structural scales contributed by classical electromagnetism ($e$), quantum mechanics ($\hbar$), and special relativity ($c$) intersect. None of these theories, taken individually, supplies the independent scales required to define $\alpha$. The constant first appears when relativistic corrections are added to the Schr\"odinger--Bohr description of hydrogen (Sommerfeld), and it becomes the structural coupling in quantum electrodynamics, where quantum and relativistic effects modify the classical electromagnetic interaction. Expressing the governing laws in canonical form reveals this dependence and eliminates representational artifacts that make $\alpha$ appear fundamental. The running of $\alpha$ in QED further demonstrates its status as a scale-dependent coupling rather than a universal constant. I conclude that $\alpha$ is a domain-specific structural ratio reflecting contingent relationships among independent physical scales.
\end{abstract}


\bigskip
\noindent\textbf{Note:} Submitted to \emph{Studies in History and Philosophy of Science}

\input{sec1-introduction}
\input{sec2-historical}
\input{sec3-scaled-quantities}
\input{sec4-emergence-alpha}
\input{sec5-canonical-forms}
\input{sec6-implications-conclusion}

\bibliographystyle{plainnat}
\bibliography{references}

\end{document}

%% file: sec1-introduction.tex
\section{Introduction}

Few quantities in physics have attracted as much metaphysical attention as the fine-structure constant,
\[
\alpha = \frac{e^{2}}{4\pi \varepsilon_{0} \hbar c} \approx \frac{1}{137}.
\]
Its numerical simplicity and dimensionlessness have encouraged the belief that $\alpha$ encodes a deep and universal feature of physical law. Richard Feynman's remark that $\alpha$ is a ``magic number'' written by an unknown hand crystallized this sentiment and helped foster the expectation that the constant should be derivable from mathematics alone \citep{Feynman1985}.

Nothing in the structure of physical theory supports this view. $\alpha$ does not appear in the fundamental equations of classical electromagnetism. It does not appear in nonrelativistic quantum mechanics. It does not appear in relativity considered independently. It emerges only when these frameworks are combined---initially in Arnold Sommerfeld's relativistic corrections to hydrogen and fully in the interaction structure of the Dirac--Maxwell theory and quantum electrodynamics (QED) \citep{Sommerfeld1916,Bethe1957}. Classical electromagnetism contributes the charge scale $e$; quantum mechanics introduces the action scale $\hbar$; and special relativity contributes the invariant speed $c$. Only where these scales coexist does the dimensionless ratio $\alpha$ become physically meaningful.

This suggests that $\alpha$ is not a fundamental constant, but a \emph{scaled quantity}. Its dimensionlessness reflects the interplay among distinct structural scales rather than mathematical fundamentality. Scaled quantities are not universal; they arise only in the domains where the relevant scales operate. The Reynolds number, the Mach number, and the plasma parameter exemplify this category, and $\alpha$ is their quantum-electrodynamic analogue \citep{Buckingham1914,Misic2010}.

Canonical representations of the governing equations make this dependence explicit. When Maxwell's equations, the Dirac equation, and the interaction terms of QED are written in forms that display all operative scales, the dependence of $\alpha$ on the joint presence of electromagnetic, quantum, and relativistic structure becomes clear. Compact notation in natural units, by contrast, conceals these scales and encourages the misconception that $\alpha$ is a pure mathematical constant. The running of $\alpha$ in QED reinforces this point: a coupling that varies with energy scale cannot be a timeless universality \citep{Peskin1995,Mohr2016}.

The conceptual puzzles surrounding $\alpha$ arise not from deep features of nature but from representational conventions and inherited assumptions about dimensions and fundamentality. When the structural roles of $e$, $\hbar$, and $c$ are made explicit, the appearance of mystery dissolves. $\alpha$ emerges as a domain-specific quantity whose value reflects the magnitudes of independent physical scales in quantum electrodynamics.

%% file: sec2-historical.tex
\section{Historical and Conceptual Background}

The modern elevation of dimensionless constants to metaphysical significance did not arise from the internal structure of physical theories. It developed through a particular intellectual and metrological history in which practical demands, representational conventions, and a preference for algebraic tidiness gradually overshadowed the physical structures that give quantities their meaning. Understanding why $\alpha$ came to be regarded as a mysterious ``pure number'' requires tracing this conceptual lineage.

\subsection{Descartes and the Birth of Algebraic Homogenization}

The genealogy begins not with physics but with algebra. In \emph{La G\'eom\'etrie}, Ren\'e Descartes introduced a symbolic framework for representing geometric magnitudes through variables such as $x$, $y$, and $z$. This extended the earlier work of Fran\c{c}ois Vi\`ete, who had also used abstract letters to denote magnitudes but insisted that algebraic operations respect dimensional homogeneity. For Vi\`ete, lines could be added only to lines, areas only to areas, and so forth.

Descartes circumvented these constraints by introducing an implicit fictitious unit length. Dividing geometric magnitudes by this arbitrary standard rendered them dimensionless, making expressions like $x^{2} + x^{3}$ algebraically permissible \citep{Descartes1637}. This transformation allowed geometrically distinct magnitudes to be manipulated symbolically within a unified algebraic framework.

The move introduced a conceptual shift. Once magnitudes were expressed relative to an arbitrary standard, numerical representation began to obscure structural distinctions. Representational convenience came to be mistaken for inherent unity. This development laid the groundwork for later assumptions that dimensionless expressions carry special physical significance even when their dimensionlessness is a product of convention rather than structure.

\subsection{Fourier, Buckingham, and the Rise of Dimensional Analysis}

By the early nineteenth century, the expansion of mathematical physics brought these issues into sharper focus. When Joseph Fourier formulated his differential equations for heat flow in \emph{Th\'eorie Analytique de la Chaleur} (1822), he introduced variables---thermal conductivity, heat capacity---that did not fit naturally within the traditional dimensional scheme of length, mass, and time. Fourier ensured dimensional consistency by assigning these quantities whatever dimensions preserved equality in his equations \citep{Fourier1822}. This was a significant early step toward a systematic dimensional framework.

A general formal treatment arrived with Edgar Buckingham's $\Pi$-theorem \citep{Buckingham1914}, which established that any physically meaningful relationship involving $n$ variables with $k$ independent dimensions can be rewritten in terms of $n-k$ dimensionless combinations. The theorem provided a principled method for reducing complex problems to a small number of unit-invariant parameters. Engineers quickly recognized its value: scale models could reliably predict full-scale behavior when the relevant dimensionless groups were matched.

The success of this method encouraged an interpretive shift. Because dimensionless groups remained unchanged under unit transformations, they came to appear physically privileged. Their invariance was often mistaken for metaphysical depth rather than representational convenience. This contributed to the later perception that dimensionless constants such as $\alpha$ reveal something uniquely fundamental about physical law.

\subsection{Electromagnetic Units and the Illusion of Fundamentality}

These tendencies became especially pronounced in electromagnetism. Electric charge does not naturally fit within the mechanical dimensions of $L$, $M$, and $T$, and a variety of unit systems emerged to manage the mismatch: electrostatic units, electromagnetic units, Gaussian units, Heaviside--Lorentz units, SI, and several unrationalized cgs schemes. Each system introduced different constants and dimensional assignments to reconcile electrical and mechanical quantities. The differences reflected representational choices, not distinct physical theories.

This proliferation of conventions encouraged the belief that the simplest or most symmetric algebraic formulation must reveal the underlying physical structure. Factors such as $4\pi$, $\varepsilon_{0}$, and $\mu_{0}$ appeared or disappeared depending purely on convention. The temptation was to interpret the ``cleanest'' formulation as the most physically faithful. Yet these differences arose from historical and metrological decisions, not from nature itself.

Jim Grozier has argued against this interpretive move. He shows that treating unit-invariant formulations as ontologically superior confuses representational tidiness with physical insight. Unit-invariance is a feature of the description, not a marker of deeper structure. Clear notation aids understanding; it does not adjudicate ontology \citep{Grozier2018}.

\subsection{Rationalization, IUPAP 1932, and the Ideal of Clean Form}

The drive toward algebraic tidiness culminated in the 1932 recommendations of the International Union of Pure and Applied Physics (IUPAP), which proposed rationalizing Maxwell's equations by eliminating geometric factors such as $4\pi$. The Commission explicitly recommended introducing a dimensional constant into Coulomb's law to separate electrical from mechanical dimensions, acknowledging that electric charge could no longer be treated as a derived mechanical quantity \citep{Griffiths1932}. This marked a major turning point in electromagnetic metrology.

Rationalization clarified which constants encode physical structure and which arise from dimensional conventions. It also reinforced the belief that the aesthetically simplest representation must be the most physically revealing. The reform reduced algebraic clutter but did not alter the underlying theory. Nonetheless, it strengthened the cultural association between dimensionless expressions and physical fundamentality.

\subsection{Compact Notation and the Suppression of Scale}

The most influential step toward the modern metaphysics of constants came with the adoption of natural units, in which foundational dimensional quantities are set equal to unity:
\[
c = \hbar = k_{B} = 1.
\]
This convention dramatically compresses notation. In natural units, $\alpha$ appears simply as $e^{2}/4\pi$, suggesting a dependence solely on charge. But this simplicity is achieved by absorbing the quantum and relativistic scales into the definitions of all remaining quantities. Time and length share identical units; energy, frequency, and inverse length become interchangeable; and distinctions among independent physical scales vanish from explicit view.

This suppression of scale encourages interpretive error. Dimensionless expressions appear to carry special physical significance when their simplicity is merely representational. Grozier's critique again applies: the algebraically simplest form need not be the most structurally revealing. The case of $\alpha$ exemplifies this. Writing $\alpha$ as $e^{2}/4\pi$ obscures its dependence on $\hbar$ and $c$ and thus masks the structural conditions under which the quantity is defined.

\subsection{Duff and the Dimensionless-Only Thesis}

A more formal expression of this attitude appears in Michael Duff's argument that only dimensionless constants can qualify as fundamental, since dimensional constants can always be eliminated through suitable unit choices \citep{Duff2002}. On this view, $c$, $\hbar$, and $k_{B}$ reflect measurement conventions, while dimensionless combinations such as $\alpha$ capture genuine physical features.

Duff's claim is correct as a point about representation but overreaches when taken as a metaphysical conclusion. Dimensionlessness does not imply fundamentality. The dimensionlessness of $\alpha$ results from the cancellation of scales, not from independence of those scales. Its definability depends on the presence of electromagnetic, quantum, and relativistic structures. Outside the domain where these structures coexist, $\alpha$ does not arise at all.

%% file: sec3-scaled-quantities.tex
\section{Scaled Quantities and Their Structural Basis}

A dimensionless ratio is often assumed to possess intrinsic significance because its numerical value does not depend on unit conventions. Yet the absence of explicit units does not by itself confer physical depth. The crucial distinction is between bare dimensionlessness and structural significance. A dimensionless ratio is meaningful only when the physical theory supplies independent scales whose comparison the ratio encodes. Without those scales, the quantity is either undefined or devoid of physical interpretation.

Even the natural numbers illustrate this point. Counting ``three electrons'' presupposes a unit of individuation---the electron as a distinguishable entity. Cardinality is unit-free only in the limited sense that the counting unit is suppressed in notation, not in the richer sense of independence from representational or physical assumptions. ``Unitless'' numbers are not therefore devoid of hidden structure; they depend on background conditions that the notation conceals. This is a caution against inferring metaphysical depth from the mere absence of dimensional factors.

Scaled quantities form a distinct and philosophically important category. They arise when a theory includes at least two independent, structurally relevant scales whose ratio encodes a physically interpretable comparison. Their dimensionlessness is a reflection of structure, not a sign of its absence.

\subsection{Reynolds Number: Dimensionless but Domain-Bound}

The Reynolds number provides a paradigmatic example:
\[
\mathrm{Re} = \frac{\rho v L}{\mu}.
\]
It is dimensionless, unit-invariant, and indispensable in fluid mechanics, yet it becomes meaningless outside the domain in which the quantities it relates are well defined. In molecular dynamics or in flows approaching the Knudsen regime, viscosity ceases to be a constitutive parameter; $\mu$ no longer exists as a state variable. When viscosity disappears, $\mathrm{Re}$ disappears with it \citep{Reynolds1883}. The significance of the ratio depends on a continuum description of the fluid, on the existence of characteristic inertial and viscous scales, and on the physical processes these scales mediate.

When $\mathrm{Re} < 10$, resistive forces scale linearly with velocity ($F \propto \eta u L$). When $\mathrm{Re} > 100$, inertial forces dominate ($F \propto \rho u^{2} S$) \citep{Misic2010}. The Reynolds number thus delineates structural regimes in which different physical processes govern behavior. It is not a universal constant but a diagnostic parameter that becomes meaningful only when viscosity, density, and characteristic length scales coexist in a theoretical description. Dimensionlessness here does not imply fundamentality; it indicates that the theory provides the structure required to form a meaningful ratio.

\subsection{The Relativistic Parameter $\beta$: A Ratio That Requires a Theory}

A similar pattern appears in the relativistic velocity parameter
\[
\beta = \frac{v}{c}.
\]
$\beta$ is formally dimensionless, but its physical meaning depends entirely on the structure of special relativity. Only in relativity does $c$ serve as a distinguished velocity scale with geometrical and dynamical significance. The ratio $v/c$ thereby acquires interpretive content: it expresses the relation between a particular velocity and the invariant limit encoded in spacetime structure.

In a Galilean framework there is no invariant speed, velocities add linearly, and the theory supplies no scale against which to normalize $v$. One may form the mathematical ratio $v/c$, but it carries no physical interpretation. $\beta$ is thus another example of a scaled quantity: meaningful only within a theory that supplies the necessary structure.

\subsection{What Scaled Quantities Are}

These examples illustrate a general profile. A scaled quantity is a dimensionless ratio $R = A/B$ where $A$ and $B$ are independent structural parameters of a theory, where the theory renders their comparison physically interpretable, and where outside that theoretical context the ratio either lacks definition or lacks meaning. The Buckingham $\Pi$-theorem formalizes part of this pattern \citep{Buckingham1914,Misic2010}, but the significance of dimensionless parameters depends on how physical structure determines the scales they relate.

Mach number, Knudsen number, plasma parameters, and many other dimensionless groups share this pattern. All are dimensionless; none are fundamental; each becomes meaningful only within domains whose structure supports the scales they compare.

\subsection{Why $\alpha$ Belongs to This Category}

The interpretive error surrounding $\alpha$ stems from ignoring the structural conditions required to form meaningful dimensionless ratios. Electromagnetism supplies a coupling scale $e$; quantum mechanics introduces the action scale $\hbar$; special relativity contributes the invariant speed $c$. Only where these independent scales coexist can $\alpha$ be defined. Its dimensionlessness does not mark it as fundamental. It is a scaled quantity whose significance depends on the structural conjunction of electromagnetic, quantum, and relativistic frameworks. Outside that conjunction, $\alpha$ is undefined.

%% file: sec4-emergence-alpha.tex
\section{The Emergence of $\alpha$ at Theory Junctions}

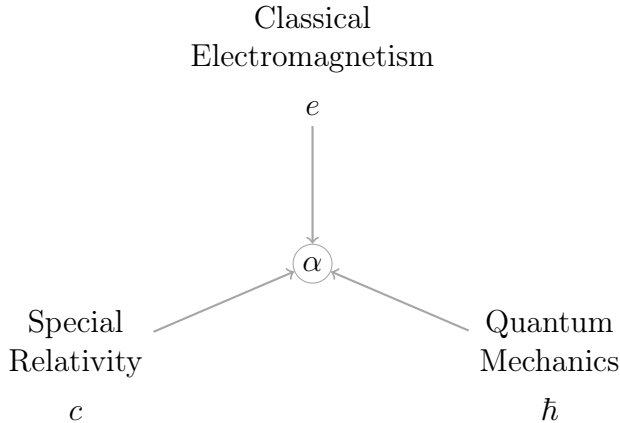
\begin{figure}[t]
  \centering
  \begin{tikzpicture}[scale=1.5]

    \node[align=center] (EM) at (0,1.8) {Classical\\Electromagnetism\\[4pt] $e$};
    \node[align=center] (QM) at (2.1,-0.9) {Quantum\\Mechanics\\[4pt] $\hbar$};
    \node[align=center] (SR) at (-2.1,-0.9) {Special\\Relativity\\[4pt] $c$};

    \node[draw=gray!70, circle, fill=white, inner sep=2pt] (alpha) at (0,0) {$\alpha$};

    \draw[->,gray!70,thick] (EM) -- (alpha);
    \draw[->,gray!70,thick] (QM) -- (alpha);
    \draw[->,gray!70,thick] (SR) -- (alpha);

  \end{tikzpicture}
  \caption{\small Schematic structural dependence of the fine-structure constant $\alpha$ on three independent frameworks. Classical electromagnetism contributes the charge scale $e$, quantum mechanics the action scale $\hbar$, and special relativity the invariant speed $c$. The arrows indicate that $\alpha$ is definable only when all three structures are jointly operative; they are not intended to represent causal or mathematical derivation.}
  \label{fig:alpha-structure}
\end{figure}

The fine-structure constant belongs to no single theory taken in isolation---not to classical electromagnetism, not to quantum mechanics, not to relativity. It arises from the intersection of theoretical frameworks that each introduce independent physical scales. What $\alpha$ reflects is how these structures combine. Figure~\ref{fig:alpha-structure} presents this schematically: classical electromagnetism, quantum mechanics, and special relativity contribute the scales $e$, $\hbar$, and $c$ respectively, and $\alpha$ is definable only when all three are jointly in play.

\subsection{Classical Electromagnetism: Scale Underdetermination and Rescaling Symmetry}

Classical electromagnetism determines how charges interact and how electromagnetic fields propagate, but it does not fix the absolute scale of electric charge. The transformation
\[
q \to \lambda q, \qquad \varepsilon_{0} \to \frac{\varepsilon_{0}}{\lambda^{2}}
\]
leaves all empirical predictions invariant. If all charges are rescaled by $\lambda$, the vacuum's response compensates proportionally, and the measurable consequences remain unchanged. This symmetry shows that Maxwell's equations contain no intrinsic electromagnetic strength scale. Charge magnitude is determined only up to an overall multiplicative constant. Classical electromagnetism provides no mechanism to predict or explain the empirical value of $e$; it takes charge as external input. For this reason, classical electromagnetism alone cannot generate a dimensionless coupling parameter that characterizes the strength of electromagnetic interaction. The theory can compute ratios using measured values, but it supplies no theoretical account of those values.

The vacuum permittivity $\varepsilon_{0}$ appears in SI formulations, but the rescaling symmetry shows that $\varepsilon_{0}$ is not an independent physical scale. It encodes how SI partitions electrical from mechanical dimensions. In Gaussian units, where the unit of charge is chosen differently, the Coulomb interaction involves $e^{2}$ directly. These representations differ in convention, not in physical content.

This situation contrasts with gravity. Newton's constant $G$ cannot be removed by a similar rescaling because gravitational charge (mass) is tied to inertial mass through the equivalence principle. Electromagnetism permits a global rescaling of charge; gravity does not. Classical electromagnetism is therefore structurally incapable of producing a dimensionless coupling such as $\alpha$.

\subsection{Nonrelativistic Quantum Mechanics: A New Scale Without a Reference}

Quantum mechanics introduces a fundamental scale: Planck's constant $\hbar$. Its presence governs interference phenomena, the spacing of energy spectra, and the structure of bound states. But $\hbar$ alone does not suffice to define $\alpha$. The ratio
\[
\frac{e^{2}}{4\pi\varepsilon_{0}\hbar}
\]
has dimensions of velocity. Without a distinguished speed against which to normalize this quantity, nonrelativistic quantum theory cannot produce a dimensionless electromagnetic coupling. Galilean invariance offers no such scale, as velocities add linearly and no invariant speed exists.

Quantum mechanics thus contributes an essential ingredient---an action scale---but cannot by itself generate $\alpha$.

\subsection{Special Relativity: A Distinguished Velocity Scale}

Special relativity introduces the invariant speed $c$, providing the missing reference scale. With $c$, the ratio $v/c$ becomes physically meaningful, and velocities acquire a structural normalization. Relativity therefore contributes the velocity scale required to render $e^{2}/\hbar$ comparable to a universal standard.

Relativity alone, however, supplies no electromagnetic scale. It provides the reference structure for forming dimensionless velocity ratios but does not determine how strong electromagnetic interactions should be. It is only when the relativistic scale $c$ is combined with the electromagnetic scale $e$ and the quantum scale $\hbar$ that $\alpha$ becomes definable.

\subsection{Emergence of $\alpha$ in Relativistic Quantum Theory}

The structural origin of $\alpha$ is clearest when these frameworks are combined. In both semiclassical and fully quantum treatments, $\alpha$ appears when electromagnetic, quantum, and relativistic scales are joined.

In 1916, Arnold Sommerfeld extended the Bohr model by incorporating relativistic corrections to the electron's motion in hydrogen. For bound electrons in the lowest orbit,
\[
\frac{v}{c} \approx \alpha,
\]
so the electron's orbital velocity at the Bohr radius is approximately $\alpha c$. The fine-structure splitting appears as a series expansion in $\alpha^{2}$. Although semiclassical, Sommerfeld's calculation identified the relativistic origin of fine-structure corrections and anticipated the full quantum treatment.

In quantum electrodynamics, the same pattern appears more systematically. Classical electromagnetism supplies the coupling parameter $e$; quantum mechanics introduces $\hbar$; relativity contributes $c$. At tree level, interaction amplitudes scale with $e$, but physical observables depend on $e^{2}$ and therefore on $\alpha = e^{2} / (4\pi \varepsilon_{0} \hbar c)$. Quantum corrections---self-energy terms, vacuum polarization, and vertex corrections---appear in powers of $\alpha$:
\[
\mathcal{O}(\alpha), \quad \mathcal{O}(\alpha^{2}), \quad \mathcal{O}(\alpha^{3}), \dots
\]
This expansion reflects the fact that $\alpha$ parametrizes the order at which quantum and relativistic corrections modify classical electromagnetism. Once the relevant scales are present and small, the theory behaves like a system organized around a controlling dimensionless parameter.

$\alpha$ thus ``crystallizes'' out of the structural conjunction of electromagnetism, quantum mechanics, and relativity. It is not an independent element of any one theory, but a ratio that becomes meaningful only when the distinct scales introduced by each framework are simultaneously operative.

\subsection{QED, Renormalization, and the Running of $\alpha$}

The dependence of $\alpha$ on structure becomes explicit in quantum electrodynamics. Electric charge is not fixed; the quantum vacuum polarizes, and the effective charge depends on the momentum scale at which it is probed. At low energies,
\[
\alpha(0) \approx \frac{1}{137.035999\ldots},
\]
while at the scale of the $Z$ boson mass,
\[
\alpha(m_{Z}) \approx \frac{1}{128.9}.
\]
The increase reflects the fact that higher-energy probes see through more of the vacuum polarization cloud and therefore experience a less screened, effectively stronger charge.

If $\alpha$ were a metaphysically fundamental constant, it would retain its character across different theoretical contexts. But the running of $\alpha$ reveals its dependence on QED's specific dynamical structure. The beta function that governs how $\alpha$ varies with scale depends on the particle content of the theory, the structure of vacuum polarization, and the details of renormalization. These are features of QED, not features of mathematics or of nature independent of theoretical framework.

A genuinely framework-independent constant---such as $\pi$ or the topological invariants of spacetime---appears with the same meaning across diverse theories. By contrast, $\alpha$ exists only within QED (and theories that reduce to it in appropriate limits), and its behavior is determined by how that specific theory implements the interaction between electromagnetic fields and charged matter. The running of $\alpha$ thus confirms its status as a structural parameter of QED rather than a universal feature of physical law.

\subsection{Where $\alpha$ Lives}

The fine-structure constant exists only at the intersection of classical electromagnetism, quantum mechanics, and special relativity. It appears explicitly in the Dirac--Maxwell Lagrangian and acquires its physical meaning through renormalization in QED. Outside this structural intersection, $\alpha$ simply does not arise.

This is illustrated by phenomena often associated with quantum electrodynamics but that do not involve $\alpha$. The Casimir effect arises from vacuum fluctuations of the electromagnetic field and depends on $\hbar$ and $c$, but it requires no coupling to charge and therefore no dependence on $e$ or $\alpha$ \citep{Casimir1948}. The quantum Hall effect yields the quantized resistance
\[
R_{K} = \frac{h}{e^{2}},
\]
which depends on $\hbar$ and $e$ but not on relativistic structure \citep{Jeckelmann2001}. Both phenomena inhabit domains where only subsets of the electromagnetic, quantum, and relativistic scales operate. $\alpha$ appears precisely when all three are present.

%% file: sec5-canonical-forms.tex
\section{Canonical Forms and Invariant Physical Structure}

Canonical forms express a theory in a way that makes all operative physical quantities explicit and separates structural content from representational convention. They do not eliminate constants; they distinguish those that encode invariant structure from those introduced by human choices of units, normalization, or notation.

\subsection{Canonical Representation as Structural Expression}

In theoretical and philosophical discussions, ``canonical'' is sometimes used loosely to mean ``standard'' or ``well formatted.'' Here the term refers to a more precise requirement: a canonical representation exposes all independent physical scales and parameters explicitly and prevents their absorption into unit definitions or normalization conventions. It displays the structure of the law independently of arbitrary representational choices.

A constant that disappears because of a unit choice has not been removed from the physics. A constant that disappears because it plays no structural role has. Confusing these two modes of disappearance has contributed to mistaken claims about fundamentality.

\subsection{Why Canonical Form Matters for $\alpha$}

Coulomb's law illustrates the point. Written canonically,
\[
F = \frac{1}{4\pi \varepsilon_{0}} \frac{q_{1} q_{2}}{r^{2}},
\]
the scale $\varepsilon_{0}$ appears explicitly. In Gaussian units the same law becomes
\[
F = \frac{q_{1} q_{2}}{r^{2}},
\]
but only because the unit of charge is defined to absorb the factor $4\pi \varepsilon_{0}$. The two forms describe the same physics, but the latter hides one of the scales that the former makes explicit. Canonical form avoids this suppression of structure.

This is especially important for understanding $\alpha$. In natural units, where $\hbar = c = 1$, $\alpha$ is written as
\[
\alpha = \frac{e^{2}}{4\pi},
\]
which suggests that the constant reflects only the magnitude of charge. But $\hbar$ and $c$ remain fully operative; their roles are merely hidden by convention. Canonical form restores the explicit dependencies:
\[
\alpha = \frac{e^{2}}{4\pi \varepsilon_{0} \hbar c}.
\]
This expression reveals that $\alpha$ requires the joint presence of electromagnetic coupling, quantum action, and relativistic invariance. The quantity depends equally on all three.

Natural units simplify calculation but can mislead interpretation. They collapse dimensional distinctions and give the impression that suppressed scales are absent from the physics. Canonical representation avoids this interpretive trap.

\subsection{The Role of $\varepsilon_{0}$ in Canonical Representation}

In SI, $\varepsilon_{0}$ appears because the dimensional scheme distinguishes electrical from mechanical quantities. The rescaling symmetry of classical electromagnetism shows that $\varepsilon_{0}$ is not an independent physical scale; it encodes how SI partitions dimensions. In Gaussian units, $\varepsilon_{0}$ is absorbed into the definition of charge. The electromagnetic coupling scale is the same in both frameworks, but its representation changes.

Canonical form keeps $\varepsilon_{0}$ explicit not to attribute independent physical significance to it, but to prevent the structural scale associated with electromagnetic interaction from being obscured by conventions.

\subsection{Canonical Form and the Visibility of Structure}

Michael Duff correctly observes that only dimensionless constants are invariant under arbitrary unit changes \citep{Duff2002}. But this fact does not imply that such constants are fundamental. A dimensionless quantity may depend on contingent structural relationships, as $\alpha$ does. Canonical representation reveals those dependencies; compact notation may conceal them.

The appearance of $\alpha$ as $e^{2}/4\pi$ in natural units suggests that it depends solely on charge. Canonical form makes clear that $\alpha$ is a ratio of three independent physical scales. Only the canonical form exhibits the full structure on which $\alpha$ depends.

\subsection{Maxwell's Equations in Canonical Form}

The covariant Maxwell equations exemplify canonical representation:
\[
\partial_{\mu} F^{\mu\nu} = \frac{1}{\varepsilon_{0}} J^{\nu}.
\]
Here the essential scales are visible. The geometrical structure of $F^{\mu\nu}$ encodes the role of $c$; $\varepsilon_{0}$ governs the coupling to matter currents; $J^{\nu}$ represents the distribution of charge and current. Variations among SI, Gaussian, and other unit systems alter the appearance of constants but not the underlying structure. Canonical form renders the invariant content explicit.

\subsection{Canonical $\alpha$: The Structural Intersection of Scales}

The canonical expression
\[
\alpha = \frac{e^{2}}{4\pi \varepsilon_{0} \hbar c}
\]
displays the three independent structural ingredients required to define the constant: electromagnetic coupling, quantum action, and relativistic invariance. None can be removed without undermining the definability of $\alpha$. The constant occupies the same conceptual category as other scaled quantities: dimensionless, unit-invariant, and structurally dependent.

\subsection{Canonical Forms and Metrology}

The 2019 redefinition of the SI fixed the numerical values of $h$ and $e$ to define human-scale units \citep{BIPM2019,Stock2019}. This change did not alter the physical constants themselves. It altered only the conventions by which units are established. In the revised SI, $\varepsilon_{0}$ becomes a derived quantity, and $\alpha$ determines how the electromagnetic coupling scale is expressed in SI units. The metrological shift aligns unit definitions more closely with canonical structure but does not elevate $\alpha$ to metaphysical fundamentality.

%% file: sec6-implications-conclusion.tex
\section{Implications for Naturalness, Structure, and the Interpretation of Constants}

The interpretation of the fine-structure constant has long been shaped by the assumption that dimensionless constants possess special metaphysical status. The foregoing analysis challenges that assumption. $\alpha$ appears not as a fundamental number written into the fabric of the world but as a contingent ratio defined within the theoretical structures that give it meaning.

\subsection{Naturalness and the Misplaced Expectation of Derivability}

The recurring effort to derive $\alpha$ from pure mathematics reflects a misunderstanding of what can legitimately be derived. In high-energy physics, ``naturalness'' concerns the stability of parameters under radiative corrections or their hierarchical relation to other scales. It does not concern the mathematical derivability of numerical values. Scaled quantities such as $\alpha$ fall outside the scope of naturalness arguments in this sense. Their values reflect the magnitudes of independent physical scales, the ways in which those scales interact, and the empirical constraints of the theories in which they appear.

Expecting $\alpha$ to emerge from pure mathematics is a category mistake. The definability of $\alpha$ requires $e$, $\hbar$, and $c$---quantities whose empirical origins are irreducible. No mathematical derivation can produce empirical magnitudes. Asking for a mathematical derivation of $\alpha$ is like asking for one of the Reynolds number: the question misidentifies the type of quantity involved.

\subsection{Framework-Dependence Without Relativism}

Framework-dependence is sometimes taken to signal arbitrariness or conventionalism. But the dependence of $\alpha$ on the joint structure of electromagnetism, quantum mechanics, and relativity does not undermine its objectivity. It indicates that $\alpha$ is meaningful only within domains that supply the scales that enter its definition.

The Reynolds number is real where viscosity, density, and characteristic length are defined; it does not become less real because it is meaningless for a dilute gas. The same holds for Mach number, Knudsen number, and other scaled parameters. $\alpha$ is similarly real within the structural domain of relativistic quantum electrodynamics. Framework-dependence marks the boundaries of applicability, not the absence of physical significance.

\subsection{Dimensionless Does Not Mean Fundamental}

Michael Duff's influential argument that only dimensionless constants can be fundamental correctly emphasizes that dimensional constants can be removed through unit choices \citep{Duff2002}. But the inference from unit-invariance to fundamentality is unwarranted. A dimensionless constant may depend on contingent structural relationships, may vary with energy scale, and may lack meaning outside the theory that defines it.

$\alpha$ exemplifies all of these features. Its dimensionlessness reflects the cancellation of electromagnetic, quantum, and relativistic scales, not independence from them. Its running under the renormalization group shows that it is not timeless. Its domain-specificity demonstrates that it is not universal. Dimensionlessness confers representational convenience, not metaphysical privilege.

\subsection{Constants as Structural Markers, Not Metaphysical Primitives}

A more illuminating taxonomy distinguishes constants by their structural roles rather than by their dimensional status. Some constants---$c$, $\hbar$, $k_{B}$---set scales that appear across many domains and are anchored in the canonical forms of broad classes of theories. Others, such as $\alpha$ or the Mach and Reynolds numbers, compare independent scales and serve as markers of how different structural elements interact in a given domain. Still others, such as $4\pi$, $\varepsilon_{0}$, and $\mu_{0}$, arise from representational conventions.

$\alpha$ belongs to the class of scaled constants. Its role is diagnostic and structural: it measures the relative magnitudes of electromagnetic coupling, quantum action, and relativistic speed. Its importance lies not in metaphysical fundamentality but in the way it organizes behavior within the theoretical domain where these scales coexist.

\subsection{Structural Pluralism and Domain-Specific Ontology}

The analysis of $\alpha$ supports a modest structural pluralism. Different physical theories introduce different scales, symmetries, and parameters. Their intersections give rise to new quantities, none of which need be reducible to a single underlying structure. Electromagnetism, quantum mechanics, and relativity each contribute distinct elements; quantum electrodynamics integrates them into a coherent framework.

$\alpha$ is real because these structures are real. Its domain-specificity reflects the pluralistic architecture of physical theory. Scientific progress often consists in uncovering new structures and identifying how they intersect, rather than in reducing all constants to a single mathematical source.

\subsection{The End of the Metaphysical Puzzle}

The appearance of $\alpha$ as a pure number in certain notational systems and its cultural elevation as a mysterious constant have encouraged the belief that it demands a mathematical explanation. But once its structural basis is recognized, the puzzle dissolves. $\alpha$ is not a Platonic constant awaiting derivation. It is a ratio of independent physical scales that coexist only within a particular theoretical domain.

Its value reflects the magnitudes of those scales, not any mathematical necessity. The search for a purely mathematical derivation of $\alpha$ is therefore misdirected. The correct explanatory target is the empirical and theoretical structure that defines the scales entering the ratio.

\section{Conclusion}

The fine-structure constant has long been regarded as a fundamental and enigmatic number, a dimensionless quantity that seems to stand apart from the physical theories in which it appears. But its actual status is quite different. $\alpha$ is a scaled quantity whose meaning depends on the presence of distinct electromagnetic, quantum, and relativistic scales. It arises only where these structures intersect, and it measures the relationship among them rather than a property that exists independently of them.

Once this structural dependence is made explicit, the expectation that $\alpha$ should be derivable from pure mathematics loses its force. Its value reflects the empirical magnitudes of the independent scales within quantum electrodynamics. Outside that structural domain, $\alpha$ simply does not exist.

Canonical forms bring this dependence into sharp relief. By making all scale-setting quantities explicit, they reveal the physical structure that underlies $\alpha$ and prevent representational conventions from obscuring its dependencies. They also show why $\alpha$ remains invariant under changes of units: its domain-specific structure is preserved across representational choices, even though its appearance may vary.

None of this diminishes the importance of $\alpha$. Its smallness organizes perturbative expansions, sets characteristic scales in atomic structure, and governs the low-energy behavior of electromagnetic interactions. But its importance is bounded by the domain in which its defining scales are operative. $\alpha$ is a structural feature of QED, not a universal constant of nature.

Reframing $\alpha$ in this way generalizes beyond this particular case. The interpretation of dimensionless constants must attend to the structural context in which they arise. Their significance derives from the scales, symmetries, and theoretical commitments that render them definable. Understanding these structures allows constants to be seen not as metaphysical primitives but as markers of how physical theories organize the world.